\documentclass[aps,prb,twocolumn,showpacs,preprintnumbers,floatfix,amsmath,superscriptaddress]{revtex4-1}
\usepackage{graphicx}
\usepackage{dcolumn}
\usepackage{subfigure}
\usepackage{wrapfig}
\usepackage{cancel}
\usepackage{color}
\usepackage{bm}
\usepackage{verbatim}
\usepackage{comment}
\usepackage{epstopdf}
\usepackage{gensymb}
\usepackage{float}

\begin{document}

\title{Tunneling spectroscopy for probing orbital anisotropy in iron pnictides}
\author{N. Plonka}
\email{Author to whom correspondence should be addressed: nachumplonka@gmail.com}
\affiliation{Department of Applied Physics, Stanford University, Stanford, California 94305, USA}
\affiliation{Stanford Institute for Materials and Energy Science, SLAC National Accelerator Laboratory, Menlo Park, California 94025, USA}
\affiliation{Geballe Laboratory for Advanced Materials, Stanford University, Stanford, California 94305, USA}
\author{A. F. Kemper} 
\affiliation{Stanford Institute for Materials and Energy Science, SLAC National Accelerator Laboratory, Menlo Park, California 94025, USA}
\affiliation{Geballe Laboratory for Advanced Materials, Stanford University, Stanford, California 94305, USA}
\affiliation{Lawrence Berkeley National Lab, 1 Cyclotron Road, Berkeley, CA 94720, USA}
\author{S. Graser} 
\affiliation{Center for Electronic Correlations and Magnetism, Institute of Physics, University of Augsburg, D-86135 Augsburg, Germany}
\author{A. P. Kampf}
\affiliation{Center for Electronic Correlations and Magnetism, Institute of Physics, University of Augsburg, D-86135 Augsburg, Germany}
\date{\today}
\author{T. P. Devereaux}
\email{Author to whom correspondence should be addressed: tpd@slac.stanford.edu}
\affiliation{Stanford Institute for Materials and Energy Science, SLAC National Accelerator Laboratory, Menlo Park, California 94025, USA}
\affiliation{Geballe Laboratory for Advanced Materials, Stanford University, Stanford, California 94305, USA} 

\begin{abstract}
Using realistic multi-orbital tight-binding Hamiltonians and the $T$-matrix formalism, we explore the effects of a non-magnetic impurity on the local density of states in Fe-based compounds.  We show that scanning tunneling spectroscopy (STS) has very specific anisotropic signatures that track the evolution of orbital splitting (OS) and antiferromagnetic gaps.  Both anisotropies exhibit two patterns that split in energy with decreasing temperature, but for OS these two patterns map onto each other under $90\degree$ rotation.  STS experiments that observe these signatures should expose the underlying magnetic and orbital order as a function of temperature across various phase transitions.
\end{abstract}

% insert suggested PACS numbers in braces on next line
%STM single particle tunneling in SC.
%Non-cuprate superconductivity: pnictides
%orbital, charge and other orders in magnetic ordering. 
%band and itinerant models of magnetic ordering

\maketitle

\section{Introduction}

High-temperature superconductivity is among the most active areas of research in condensed matter physics, with most of the focus on the initially discovered cuprates \cite{Bednorz1986} and the newer iron-based superconductors. \cite{Kamihara2008}  Both of these materials also possess intriguing normal states, involving magnetic order and electronic anisotropy, which have received much attention. \cite{ DelaCruz2008,  Si2008,Fang2008,Yildirim2008,Kruger2009,Chen2010,Fernandes2012b,Kemper2009,Chuang2010,Hoffman2011,Yi2011,Kasahara2012}  Unlike cuprates, the iron materials require multiband models since all iron d-electron orbitals contribute to the low-energy physics.  As such, there are many important facets of orbital effects and their interplay with magnetic and structural order across the rich phase diagram in the pnictides.

%next 2 pgphs corr to III.A,C
Structural, spin, and orbital anisotropy are observed at various temperatures.  Neutron scattering displayed a collinear antiferromagnet (AFM) and an orthorhombic structural distortion, \cite{DelaCruz2008} which have similar transition temperatures, $T_{\mathrm{AFM}}$ and $T_\mathrm{s}$.   Many experiments have observed electronic anisotropy \cite{Chu2010a,He2010,Tanatar2010,Dusza2011,Harriger2011,Kim2011,Kuo2011,Song2011,Yi2011,Arham2012,Dusza2012,Jesche2012,Kasahara2012,Yi2012,Zhang2012} even as high as 30~K above $T_{\mathrm{AFM}}$ and $T_\mathrm{s}$. \cite{Kasahara2012}   Specifically, orbitally split $d_{xz}$- and $d_{yz}$-dominated bands have been observed in both BaFe$_2$As$_2$ and NaFeAs. \cite{Yi2011,Yi2012}  There seems to be cooperation among orbital, structural, and magnetic anisotropies since they all decrease in magnitude with doping.  However, their microscopic source needs to be further clarified. \cite{Si2008,Fang2008,Yildirim2008,Fernandes2012b,Xu2008,Haule2008,Mazin2008a,Yin2008,Han2009,Wysocki2011,Stanek2011,Kamiya2011,Daghofer2012a,Daghofer2012,Kruger2009,Singh2009,Chen2009,Lv2009,Lee2009a,Kubo2009,Chen2010,Lv2010,Bascones2010,Yanagi2010,Kontani2011,Lv2011,Nevidomskyy2011,Applegate2012,Cvetkovic2009,Knolle2010a,Brydon2009,Zhai2009,Fernandes2012b}

Understanding the microscopic Hamiltonian would shed light on superconducting and normal state properties.  For example, inter-orbital electron-electron interactions, possibly driven by strong electron-phonon coupling, yield $s_{++}$-wave superconductivity from orbital fluctuations.\cite{Kontani2010, Yanagi2010a}  These naturally give rise to orbital ordering (OO).  Alternatively, strong intra-orbital interactions, driven by spin fluctuations, produce $s_\pm$- or $d$-wave superconductivity.\cite{Kuroki2008,Mazin2008b,Parish2008,Seo2008,Chen2009a,Wang2009}  This naturally results in anisotropic spin-density waves (SDW) from nesting in the parent compound.  Such interactions may also lead to OS, \cite{Bascones2010}  for example via nematic spin fluctuations. \cite{Daghofer2012a,Daghofer2012,Fernandes2012b}  This view may be supported if the OS is much smaller at the Brillouin zone (BZ) center than the zone boundary, whereas a similar OS throughout would allow for ferroorbital ordering as the cause. \cite{Daghofer2012a,Daghofer2012}  However, the presence of zone center OS is unclear because the relevant bands are above the Fermi energy, $E_F$, which ARPES does not image.   An alternative study is therefore needed.
%Daghofer shows spin fluctuations not cause OS by Gamma, but Fernandes's analysis couples ferroorbital ordering (i.e. even OS at gamma) with spin nematic.  Therefore, I cannot definitively associate lack of OS by Gamma with spin fluctuations.

%Song et al is actually (STS) below $T_s$, but in SC part of PD; no AFM to cause nematicity.  Argues against struc causing b/c distortion so small
STS has been pivotal in imaging anisotropic patterns around impurities in the local density of states (LDOS) and in its Fourier transform, attributed to quasiparticle interference (QPI). \cite{Chuang2010,Zhou2011,Song2011,Machida2012,Allan2013,Rosenthal}  In the pnictides, these features may be attributed to a number of different sources, including itinerant \cite{Kemper2009,Knolle2010,Huang2011,Rosenthal} or local \cite{Mazin2011} anisotropic antiferromagnetism, OO, \cite{Inoue2012, Kang2012a}, anisotropic potentials, \cite{Allan2013} or impurity-pinned magnetic order. \cite{Gastiasoro}   All involve $C4 \rightarrow C2$ symmetry breaking by distinguishing between $x$ and $y$ directions. Based on the STS experiments to date, it remains a challenge to clearly identify specific features of antiferromagnetism and OO.

% Kang has quantitative agreement, and I don't see any problems, unlike what Hoffman points out in others (although Kang has yet to show QPI).  However, I doubt the community is buying into PoDW's any time soon

In the present study, we propose using STS-derived LDOS patterns as a way to distinguish OS and AFM.  As both produce band extrema (BE), i.e. local maxima or minima in the bands, that break $C4$ symmetry, $C4$ broken LDOS patterns at these BE differentiate between OS and AFM.  For OS, the difference between BE energies identified by STS patterns is the OS magnitude.  Its presence or absence above $T_\mathrm{s}$ would address the connection between OS and orthorhombicity.  Our theory employs a realistic five-orbital model with a self-consistently derived mean field SDW order parameter.  We provide prescriptions regarding energy and temperature ranges in which to find specific signatures. 

The structure of this paper is as follows.  In Sec. \ref{sec:theory}, we present the five-orbital model, including a simple phenomenological OS and a mean field orbitally resolved SDW, and outline the $T$-matrix theory used to generate tunneling spectra.  In Sec. \ref{sec:results}, we present STS patterns that track the OS magnitude via BE splitting and explain their connection to the OS dependence on magnetism (Sec. \ref{sec:OSmigration}).  In Sec. \ref{sec:SDWmigration}, we show how BE splitting also arises due to AFM gaps and how it can be distinguished from OS.  In Sec. \ref{sec:orbspin}, we present further STS behavior that can be attributed to the interdependence of magnetic and orbital ordering.  Finally, we summarize and outline the experimental steps needed to discover these signatures in Sec. \ref{sec:conclusion}.

\section{Theory}
\label{sec:theory}

We develop theory to obtain OS and SDW effects due to BE splitting, which occurs in all observed iron pnictides.  We present results for a five orbital model of LaFeAsO, \cite{Graser2009} but the five Fe $d$ orbitals are key for all pnictides and their low energy band structures are qualitatively similar.  In particular, they all have $d_{xz}$ and $d_{yz}$ BE at high symmetry points in the BZ that exhibit OS. \cite{Chen2010,Yi2011,Yi2012,Lv2011,Deng2009,Kusakabe2009}  They also may have AFM gaps that give rise to BE and need to be distinguished from those due to OS.  Although they exhibit different OS and AFM strengths, we show that the resulting BE splittings yield similar results.  We make detailed choices for material parameters and OS and AFM magnitude, but the resulting BE splitting effects are general to pnictides.

We employ five-orbital models with the kinetic energy part,
\begin{equation}
H_0=\sum_{{\bf k}\sigma}\sum_{rs} (\varepsilon_{\bf k})_{rs} c^\dagger_{{\bf k}r\sigma}c_{{\bf k}s\sigma} .
\label{H0}
\end{equation}  
Here $c^\dagger_{{\bf k}r\sigma}$ creates an electron with momentum ${\bf k}$ and spin $\sigma$ in the orbital $r$ ($s$ is also an orbital index) and $(\varepsilon_{\bf k})_{rs}=\xi_{rs}({\bf k})+(\epsilon_r-\mu)\delta_{rs}$.  Hopping parameters $\xi_{rs}({\bf k})$ and the orbital energies $\epsilon_r$ are listed in the appendix of Ref. \onlinecite{Graser2009}.  This part of the model results from a tight-binding fit to the density-functional theory band structure of LaFeAsO, for which two dimensional bands are  sufficient. \cite{Graser2009} 

Unless otherwise noted, the chemical potential $\mu$ is adjusted to maintain a filling of $n_\mathrm{tot}=6$ electrons per unit cell corresponding to undoped stochiometric pnictide materials.  An orbital basis is chosen that is aligned parallel to the nearest neighbor Fe--Fe direction and hence we use a 1 Fe per unit cell BZ.  We renormalize bands down by a factor $3$, so that the energies are closer to those observed with photoemission in BaFe$_2$As$_2$ and NaFeAs. \cite{Yi2011,Yi2012}  Throughout, we set lattice constants $a=1$ and use meV for energy units.

We analyze onsite multiorbital electron-electron interactions
\begin{eqnarray}
H_\mathrm{int} &=& U\sum_{i}\sum_{r}n_{i,r\uparrow}n_{i,r\downarrow} + U'\sum_{i}
\sum_{r,s<r}\sum_{\sigma,\sigma'}n_{i,r\sigma}n_{i,s\sigma'} \nonumber \\
& & +J\sum_{i}\sum_{r,s<r}\sum_{\sigma,\sigma'}c_{i,r\sigma}^{\dagger}
c_{i,s\sigma'}^{\dagger}c_{i,r\sigma'}c_{i,s\sigma}  \label{eq:Hint} \\
& &+J'\sum_{i}\sum_{r,s<r}\left(c_{i,r\uparrow}^{\dagger}c_{i,r\downarrow}^{\dagger}
c_{i,s\downarrow}c_{i,s\uparrow}+\mathrm{h.c.}\right) \nonumber 
\end{eqnarray}
where $i$ denotes the lattice site and $n_{i,r}=n_{i,r\uparrow}+n_{i,r\downarrow}$.
Spin rotational invariance requires $J'=J$ and in the absence of exchange anisotropy $U'=U-2J$ holds for each pair of interacting orbitals. \cite{Oles1984}  We apply these simplifications to our Hamiltonian.

\begin{figure}  
\begin{center}
\includegraphics[scale=0.6, viewport=0.25in 1.5in 7in 7.15in]{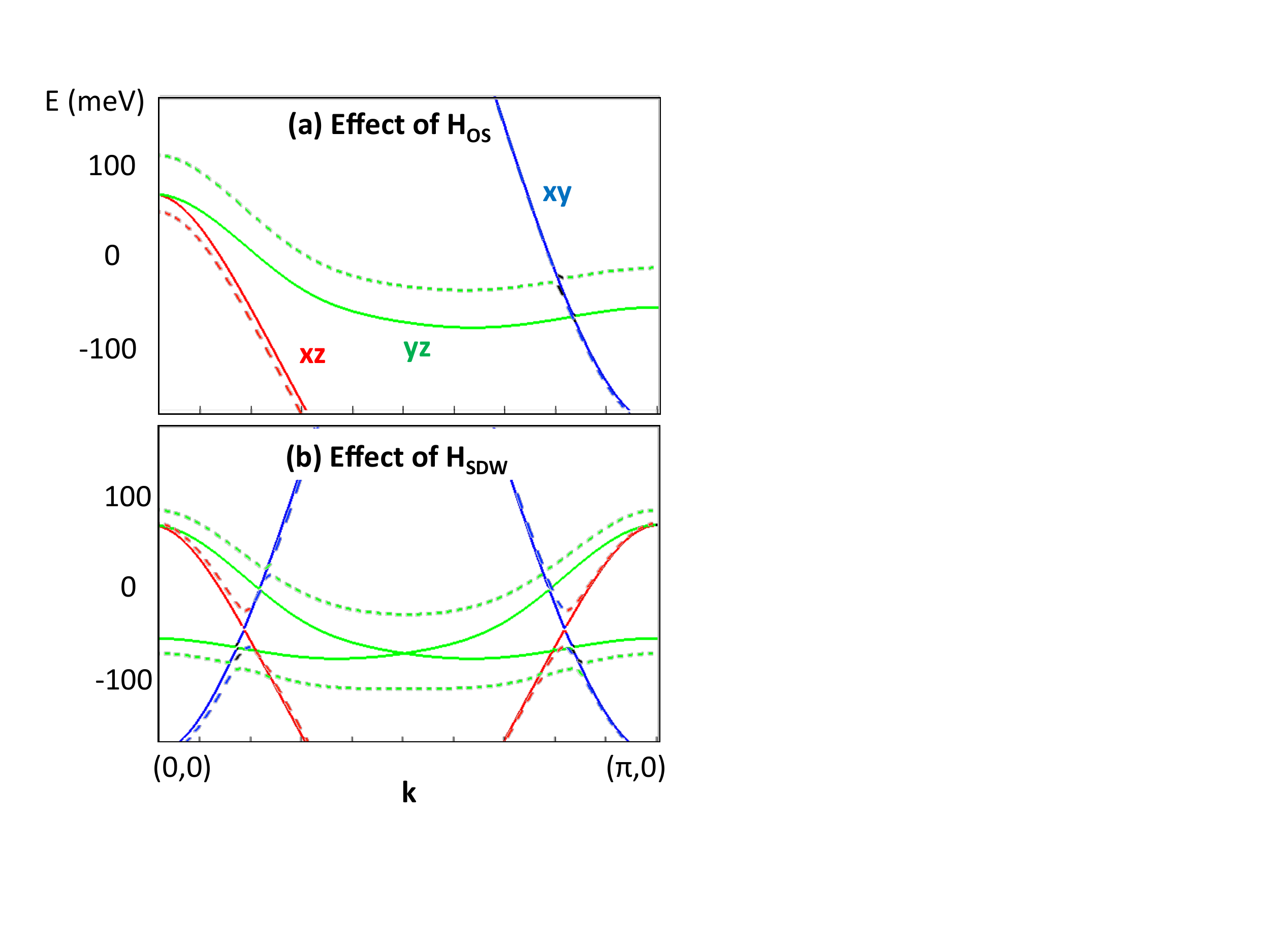}
\end{center}
\caption{(Color online) Bands along one high-symmetry direction, colored according to dominant (i.e. $>50\%$) orbital content, showing the effects of the symmetry breaking Hamiltonian terms. (a) The solid line results from the tight-binding Hamiltonian $H_0$, and the dashed and dotted lines show the OS due to $H=H_0 + H_{\mathrm{OS}}$.  (b) The effect of $H_{\mathrm{SDW}}$ is to fold and gap the bands.  The solid line is $H_0$ folded by $(\pi,0)$ into the magnetic BZ, and the dashed and dotted lines are from $H_0+H_{\mathrm{SDW}}$.  }
\label{fig:bands}
\end{figure}

To model the materials at $T_{\mathrm{OS}}>T>T_{\mathrm{SDW}}$, we explicitly introduce OS into the Hamiltonian.  Its presence is empirically verified by experiment and both electron-electron and electron-phonon interactions may naturally give rise to OO. \cite{Kruger2009, Bascones2010,Saito2010}  Since we are interested primarily in the STS phenomenology, we keep the OS fixed \cite{Yi2011,Lv2011}
\begin{align}
&H=H_0 + H_{\mathrm{OS}} , \\
&H_{\mathrm{OS}}= \sum_{{\bf k}\sigma} \Delta_{\mathrm{OS}} \left(\frac{3}{4} c^\dagger_{{\bf k},yz,\sigma}c_{{\bf k},yz,\sigma} - \frac{1}{4} c^\dagger_{{\bf k},xz,\sigma}c_{{\bf k},xz,\sigma}\right) 
\label{Hos}.
\end{align}
The OS magnitude, $\Delta_{\mathrm{OS}}$, is the energy splitting between $d_{xz}$ and $d_{yz}$ bands, and its monotonic increase reflects decreasing temperature. \cite{Yi2011,Yi2012,Kim2011}  The $3:1$ ratio and the maximum $\Delta_{\mathrm{OS}}=60$~meV that we use follow the observed quantities.\cite{Yi2011}   $H_{\mathrm{OS}}$ shifts the bands down (up) according to their amount of $d_{xz}$ ($d_{yz}$) content, as in Fig. \ref{fig:bands}(a).  The key feature is the OS at BE, whereas the full momentum dependence of the putative OS term is not as important for the small band changes considered.

For $T<T_{\mathrm{SDW}}$, we perform a mean-field decoupling of the interaction in Eq. \ref{eq:Hint}, employing the SDW order parameters
\begin{equation}
m_{rs}^{\bf Q} =\left\langle{1\over N} \sum_{i} e^{i{\bf Q}\cdot{\bf r}_i} m_{i,rs} 
\right\rangle ,
\label{mQrs1}
\end{equation}
where we take ${\bf Q}=(\pi,0)$ and the elements of the magnetization matrix at site $i$ are
\begin{equation}
 m_{i,rs} = \sum_{\sigma} \sigma c_{i,r\sigma}^{\dagger}c_{i,s\sigma} \, .
\label{mirs}
\end{equation}
The complete mean-field decoupling of the interaction term Eq.~\ref{eq:Hint} with respect to the local SDW order parameters leads to
\begin{equation}
H_\mathrm{SDW} = -\sum_{i}\sum_{r,s,\sigma}\sigma 
c_{i,r\sigma}^{\dagger}
(M_{i})_{rs}c_{i,s\sigma}.
\label{HMFint}
\end{equation}
The matrix $(M_{i})_{rs}$ is given by
\begin{equation}
(M_{i})_{rs}=\frac{J}{2}\left\langle m_{i,\mathrm{tot}}\right\rangle
\delta_{rs}+\left(\frac{U}{2}-J\right)\left\langle m_{i,sr}\right\rangle 
+\frac{J}{2}\left\langle m_{i,rs}\right\rangle \, .
\label{Umirsrot}
\end{equation}
Here we have introduced the sum of the orbitally projected moments, $m_{i,\mathrm{tot}} = \sum_r m_{i,rr}$.  

\begin{figure}
\begin{center}
\includegraphics[width=\columnwidth, viewport=0in 1in 10in 7.15in]{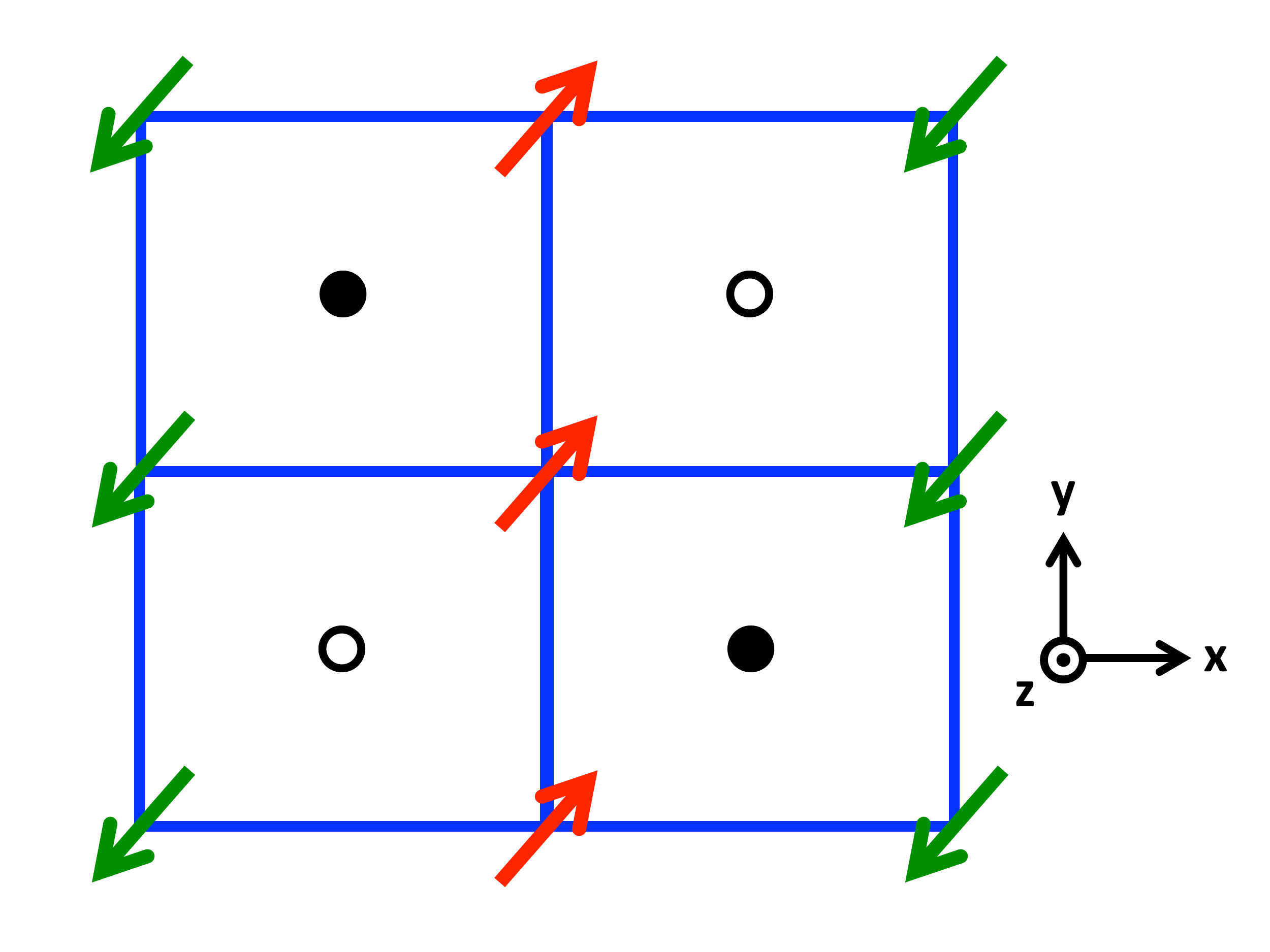}
\end{center}
\caption{(Color online) Illustration of the symmetry aspects of a magnetic FeAs
layer with $(\pi,0)$ SDW order.  Arrows mark the positions of Fe atoms with spin projection up or down. White and black circles indicate the positions of the As atoms, which are either located above or below the plane of Fe atoms.}
\label{fig:FeAslayer}
\end{figure}

$(M_{i})_{rs}$ is mostly diagonal, except for the two possible nonzero off-diagonal entries in the $e_g$ sector, as required by symmetry arguments. \cite{Lv2011}   We imagine a single FeAs layer in the "striped" $(\pi,0)$ state, as schematically
drawn in Fig. \ref{fig:FeAslayer}, and consider the following symmetry 
operations: reflection with respect to the vertical y-axis
plus inversion with respect to the perpendicular z-axis. This transformation 
leaves the spin structure of the $(\pi,0)$ state unchanged. However, the 
operator ${\hat A}=c^\dagger_{i,xz,\sigma}c_{j,yz,\sigma}$ changes into 
$-{\hat A}$. If we define the unitary operator by $R_a$, which 
corresponds to the above described symmetry operation, then obviously 
$R_a^+{\hat A}R_a=-{\hat A}$. This implies necessarily $\langle{\hat A}
\rangle=0$ and therefore $m^{\bf Q}_{rs}=0$ for $s\equiv d_{xz}$ and $r\equiv
d_{yz}$.  Conversely, for ${\hat B}=c^\dagger_{i,x^2-y^2,\sigma}c_{j,3z^2-r^2,\sigma}$ the 
symmetry operation yields $R^+_a{\hat B}R_a={\hat B}$ and hence the expectation
value of the latter operator may indeed be finite in the $(\pi,0)$ SDW state.  Similar arguments with various reflection axes require all other off-diagonal components to be zero.  

The translationally invariant density matrix is
\begin{equation}
n_{rs} = \left\langle {1\over N}\sum_{{\bf k},\sigma}\sigma 
c_{{\bf k},r\sigma}^{\dagger}c_{{\bf k},s\sigma} \right\rangle ,
\end{equation}
in which $n_{rr}$ denotes the orbital occupation per site for orbital $r$.  
$(M_{i})_{rs}$, $m_{rs}^{\bf Q}$, $\mu$, and $n_{rs}$ are determined self-consistently for a given temperature and the fixed total filling.  This involves the diagonalization of the Hamiltonian in $10 \times 10$ matrix form.  Its block form in momentum space is
\begin{align}
{\hat {\mathcal H}(k)} &= \left( \begin{array}{cc}
                       \langle {\mathbf k}|\hat{\mathcal H}|{\mathbf k}\rangle  & \langle {\mathbf k}|\hat{\mathcal H}|\mathbf{k+Q}\rangle  \\
                       \langle \mathbf{k+Q}|\hat{\mathcal H}|{\mathbf k}\rangle  & \langle \mathbf{k+Q}|\hat{\mathcal H}|\mathbf{k+Q}\rangle  \\
                      \end{array} \right) \notag  \\
           			&= \left( \begin{array}{cc}
                       \hat{\varepsilon}_{k} & \hat{M} \\
                       \hat{M} & \hat{\varepsilon}_{k+Q} \\
                      \end{array} \right) ,
\label{eq:diagH}
\end{align}
where the matrix ${\hat\varepsilon}_{k}$ contains the kinetic energy entries $(\varepsilon_{\bf k})_{rs}$ and the entries of $\hat{M}$ are $M_{rs} ={1\over N} \sum_{i} e^{i{\bf Q}\cdot{\bf r}_i} (M_{i})_{rs}$.  

In the following we report results for $U=433$~meV and $J=0.25\,U$, taking the view that intra-orbital repulsion dominates.  This gives a small zero-temperature moment $m_{i,\mathrm{tot}}=0.43$, \cite{Daghofer2010} as expected for this material. \cite{DelaCruz2008}  Furthermore, in self-consistent calculations of OS, \cite{Bascones2010, Daghofer2010} similar parameter choices result in significant OS.  Daghofer \emph{et al.} \cite{Daghofer2010} argue the OS is small, since the orbital polarization per site $n_{xz} - n_{yz}$ of order 0.1 is much smaller than the magnetic moment.  However, we found this polarization corresponds to an OS of around 60~meV in the band structure, as in experiments. \cite{Yi2011,Yi2012}  This is similar to the SDW gap magnitudes, as in Fig. \ref{fig:bands}.

\begin{figure}
\includegraphics[scale=0.6, viewport=0.75in 1in 7in 8in]{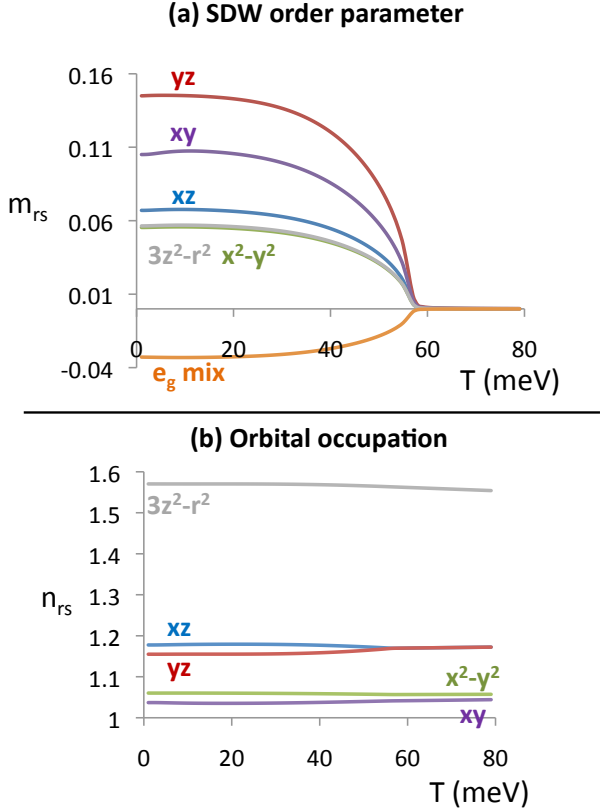}
\caption{(Color online) (a) The non-zero components of the mean-field SDW order parameter matrix $m^{\bf Q}_{rs}$ vs. temperature, for $U=433$~meV and $J=0.25\,U$.  There is only one off-diagonal component ``$e_g$ mix'', for which $r=x^2-y^2, s=3z^2-r^2$.  The \emph{diagonal} $x^2-y^2,3z^2-r^2$ components are indistinguishably close. (b) The diagonal components of the density matrix, equivalent to orbital occupation.  The nonzero term between $e_g$ orbitals is small and not shown.  A small asymmetric occupation of $d_{xz}$ and $d_{yz}$ appears with the onset of SDW order.}
\label{fig:mij}
\end{figure}

The strong intra-orbital repulsion intimately couples both OS and SDW magnetism. The orbital dependence is shown for the projected moments in Fig. \ref{fig:mij}(a).  We use a momentum mesh of $300 \times 300$ points.  Most nonzero entries of $(M_{i})_{rs}$ are on the diagonal, which corresponds to intra-orbital gaps, i.e. between bands with similar orbital content. Therefore, $H_{\mathrm{SDW}}$ gaps band crossings according to the orbital similarity between the crossed bands, as in Fig. \ref{fig:bands}(b).  A gap opens in the crossing of $d_{yz}$-dominated bands, as well as the crossing of the $d_{xz}$ and $d_{xy}$ bands, since they have similar subdominant orbitals.  Conversely, the crossing of $d_{yz}$ and $d_{xy}$ bands does not open a gap because the bands have no orbital similarity.  Notice that as the gaps open, the bands shift and cause a small OS between $d_{xz}$ and $d_{yz}$ bands, even without including $H_\mathrm{OS}$.  This affects the whole BZ, as seen from the splitting in the orbital occupations in Fig. \ref{fig:mij}(b).

We model temperatures $T<T_{\mathrm{SDW}}$ with not only the SDW term but also OS: $H = H_{0} + H_{\mathrm{OS}} + H_{\mathrm{SDW}}$.  We allow OS to affect the SDW by performing the self-consistent calculation as before but with holding $\Delta_{\mathrm{OS}}=60$~meV fixed throughout the calculation.  Although SDW alone causes some OS, we include $H_{\mathrm{OS}}$ because the presence of OS above $T_{\mathrm{SDW}}$ suggests that there is an SDW-independent contribution.  Also, the OS we obtain due to SDW alone is less than $10$~meV at $\mathbf{k}=(\pi,0)$ down to $T \ll T_{\mathrm{SDW}}$, i.e. much smaller than measured.
% The implicit admission here is that ARPES alone has showed us that there is some OS independent of SDW, and I am just not sure how much.  In contrast, reading end III.A. may give you the impression that everything is up for grabs.  No problem, intro is fairly open about what we know.

In order to investigate impurity effects, we employ the $T$-matrix for a single non-magnetic impurity, using matrix Green function methods. \cite{Han2009a,Balatsky2006,Andersen2003,Hirschfeld1988}  We start with the $10 \times 10$ Hamiltonian matrix in Eq. \ref{eq:diagH} for the SDW phase or its upper left $5 \times 5$ block in the absence of SDW.  The bare Green function, without the impurity, is $\hat{G}_0({\mathbf k},\sigma,\omega)=(\omega \hat{\mathbf 1} - \hat{\mathcal H}({\mathbf k},\sigma))^{-1}$.  We assume  momentum-independent impurity scattering, $\hat{T}(\mathbf{k,k'};\sigma;\omega) = \hat{T}(\sigma;\omega)$, as arises from a site energy shift by $V_0$ at the impurity's position.  
\begin{align}
\hat{T}(\sigma,\omega) 	&= [\hat{\mathbf 1}-\hat{V} \hat{g}_0(\sigma,\omega)]^{-1} \hat{V}, \label{eqn:T} \\
		\hat{g}_0(\sigma,\omega) &\equiv (1/N)\sum\limits_{\mathbf k} \hat{G_0}({\mathbf k},\sigma,\omega), \\
\hat{V} 	&= V_0 \left( \begin{array}{cc}
                       \hat{\mathbf 1} & \hat{\mathbf 1}  \\
                       \hat{\mathbf 1} & \hat{\mathbf 1}  \\
                      \end{array} \right).
\end{align}
There are a range of impurity strengths expected among the pnictides.  \cite{Nakamura2011,Berlijn2012} We assume a non-magnetic scattering strength $V_0 = 300$~meV for all orbitals, consistent with density-functional calculations on Co-doped BaFe$_2$As$_2$. \cite{Kemper2009}  While the qualitative results presented here are clearest near this impurity strength, they are still obtained for strengths between $100$ to $600$~meV. Hopping disorder also leads to qualitatively similar results. \cite{Vishik2009}  Note that only the diagonal entries of each block in $\hat{V}$ are non-zero, so the scattering is intra-orbital.  We avoid inter-orbital scattering, \cite{Zhang2009} as it violates reflection symmetries about the $x$ and $y$ axes and leads to unphysical LDOS patterns.  The full Green function at position $\mathbf{r_0}$ is
\begin{align}
\hat{G}(\mathbf{r_0,r_0},\sigma,\omega) &= \hat{G}_0(\mathbf{r_0,r_0},\sigma,\omega) \,+ \notag \\
&\hat{G}_0(\mathbf{r_0,0},\sigma,\omega) \hat{T}(\sigma,\omega) \hat{G}_0(\mathbf{0,r_0},\sigma,\omega), \label{eq:fullG}
\end{align}
where $\hat{G}_0(\mathbf{r,r'},\sigma,\omega)$ is the Fourier transform of $\hat{G}_0({\mathbf k},\sigma,\omega)$.  Then, the LDOS contributions from a single orbital $s$ is 
\begin{align}
\mathrm{LDOS}(\mathbf{r_0},\sigma,\omega,s) &= (-1/\pi) \mathrm{Im} [G_{ss}(\mathbf{r_0,r_0};\sigma;\omega+i\delta)] \nonumber \\
	&= \mathrm{DOS}(\sigma,\omega,s) + \delta n(\mathbf{r_0},\sigma,\omega,s),
\label{eq:LDOS}
\end{align}
with $\hat{G}_{ss}$, a $2 \times 2$ matrix with SDW or a scalar without.  The two terms in Eq. \ref{eq:LDOS} are the spatially uniform DOS and the impurity-contributed LDOS.  The total LDOS without orbital resolution is the sum of the 5 orbital LDOS's.    To calculate the QPI patterns in momentum space, either the full or the orbitally projected LDOS is Fourier transformed.  We set $\delta=3$~meV to resolve STS patterns that are split by a few meV.  We used a $100 \times 100$ real space grid with closed boundary conditions, where impurity effects are negligible.

\section{Results and Discussion}
\label{sec:results}

\subsection {Signatures of Orbital Splitting}
\label{sec:OSmigration}

\begin{figure*}
\begin{center}
\includegraphics[scale=0.65, viewport=0 1.3in 10in 8in]{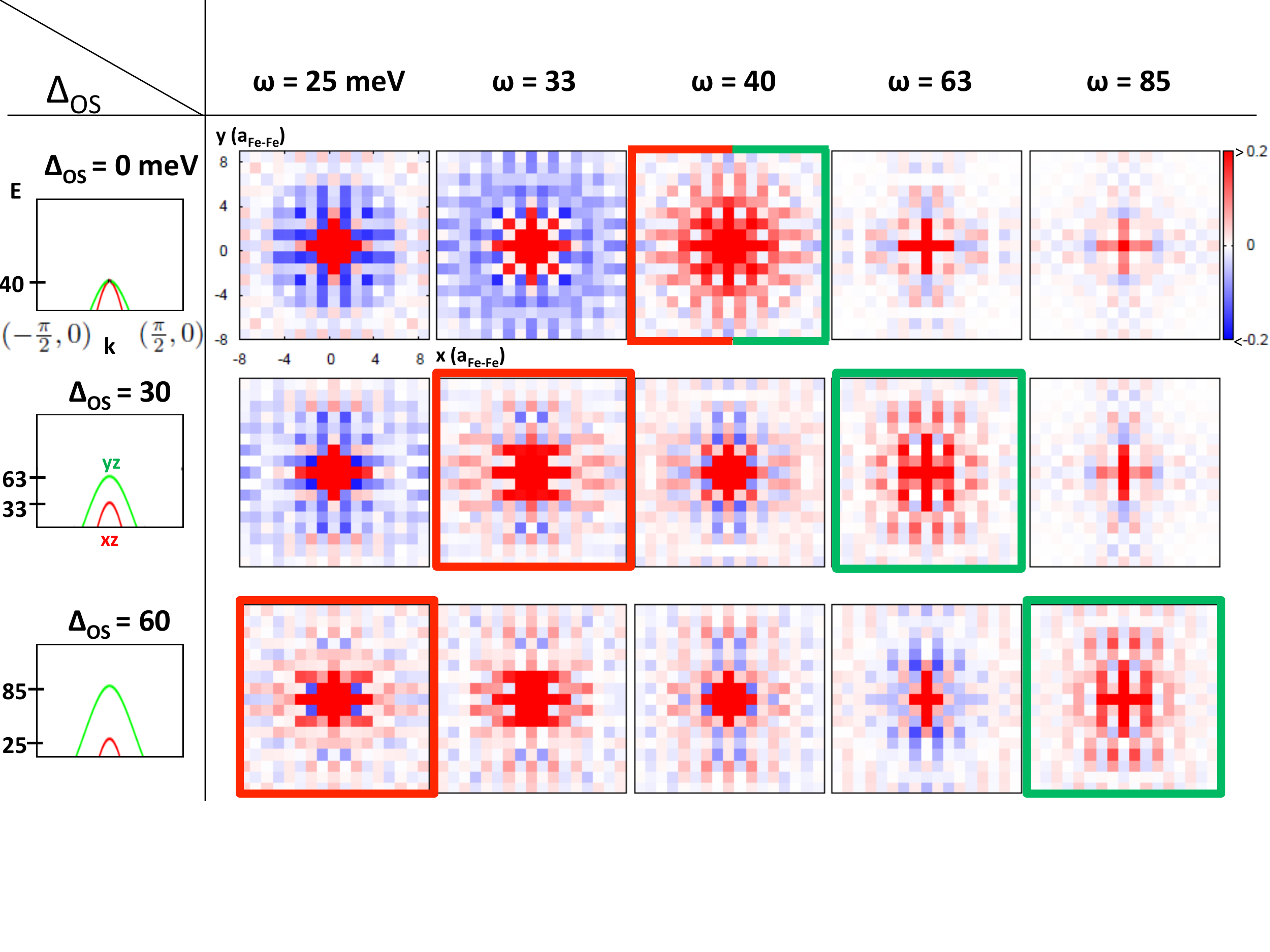}
\end{center}
\caption{(Color online) Signatures of OS in LDOS spectra due to an impurity.  Each panel shows $\delta n$ around an impurity at the origin, for a given energy $\omega$ and OS magnitude $\Delta_{\mathrm{OS}}$.  The panels are organized into rows and columns according to OS and energy, respectively.  As the left column shows, each row displays results for increasing OS, as bands of $d_{xz}$ and $d_{yz}$ content (displayed here along the $k_x$ axis) split increasingly in energy.  The LDOS patterns for each $\Delta_{\mathrm{OS}}$ outlined in red (green) are at the $\mathbf k=(0,0)$ energy of the $d_{xz}$ ($d_{yz}$) band for the selected $\Delta_{\mathrm{OS}}$.  These highlighted patterns are the signatures of OS in the LDOS, as explained in the text.}
\label{fig:OSmig_LDOS}
\end{figure*}

\begin{figure*}  
\begin{center}
\includegraphics[scale=0.65, viewport=0 1in 12in 8in]{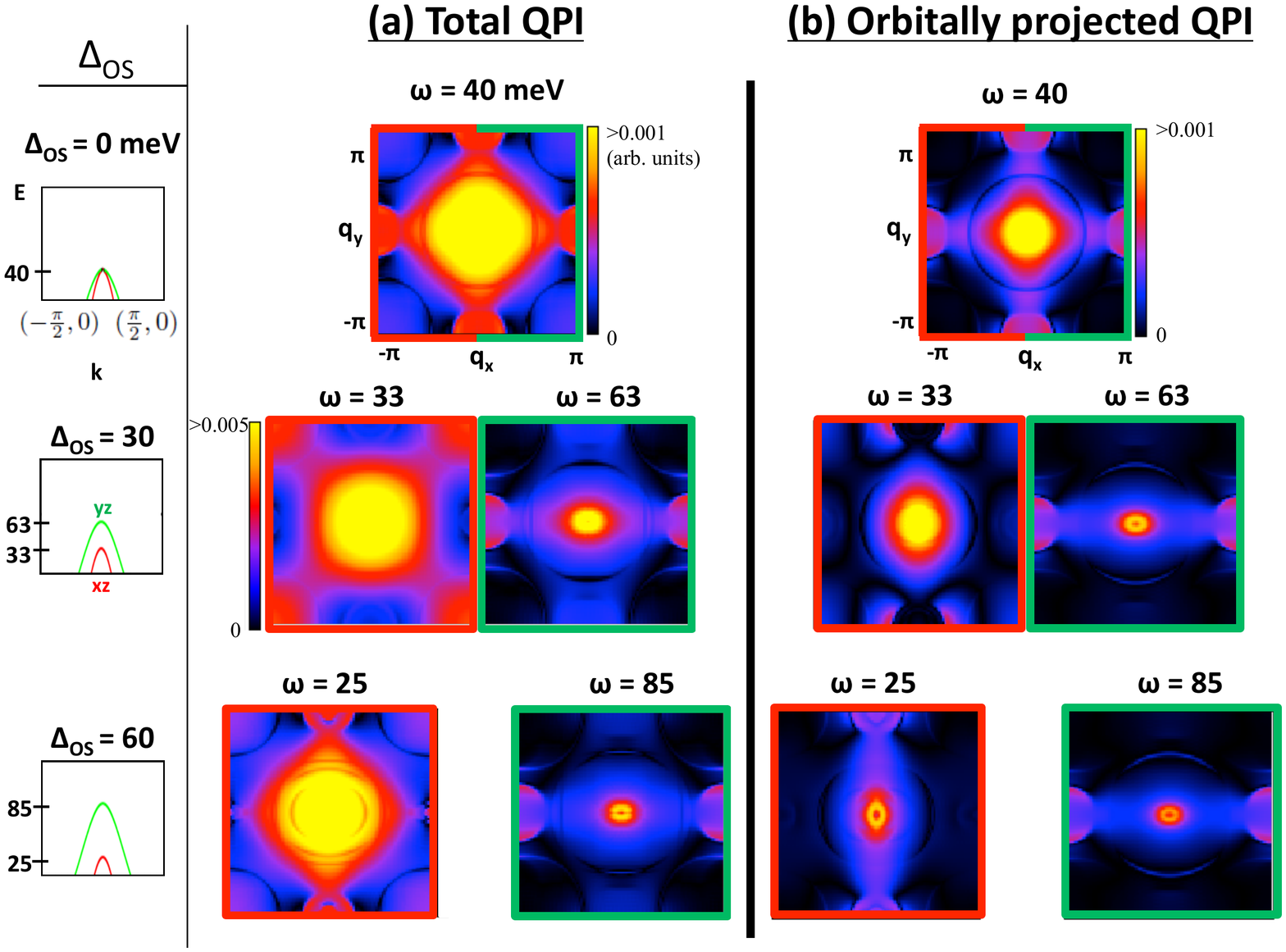}
\end{center}
\caption{(Color online) (a) Total QPI patterns (absolute value), i.e. the Fourier transform of $\delta n$, for the same energies and OS as the \emph{outlined} LDOS's in Fig. \ref{fig:OSmig_LDOS}.  Here, the OS signatures are partly obscured.  All QPI patterns have the same intensity scale, from 0 to 0.001, except at $\omega=33$~meV.  (b)  Orbitally projected QPI, which displays OS signatures.  For $\omega < 40$~meV ($\omega > 40$~meV), only the QPI contributions from $d_{xz}$ ($d_{yz}$) orbitals are shown, and at $\omega = 40$~meV, we show the absolute value of the sum of $d_{xz}$ and $d_{yz}$ contributions. All patterns have the same intensity scale.}
\label{fig:OSmig_qpi}
\end{figure*}

We present STS patterns to identify OS signatures involving BE at two nearby energies with increasing splitting as temperature decreases.  While these are clearly resolved in LDOS spectra they are only partially revealed in QPI.  The distance between two BE energies is a measure of the magnitude of OS.  We explain how a comparison to the SDW gap magnitudes clarifies the OS dependence on magnetism.

The LDOS identifies and tracks OS by BE splitting due to a large contribution to $\delta n$ from the flat portions of the band structure.  In Fig. \ref{fig:OSmig_LDOS}, we plot LDOS patterns due to an impurity with $H = H_{0} + H_{\mathrm{OS}}$.  We show the LDOS evolution for several energies as temperature decreases, mimicked by an increase in $\Delta_{\mathrm{OS}}$ in the figure's left column.   Patterns with finite $\Delta_{\mathrm{OS}}$ exhibit $C4 \rightarrow C2$ symmetry breaking, but more detailed signatures are needed to distinguish OS from other sources of $C4$ symmetry breaking such as the SDW.  Just above $E_F$ at the zone center, there are two BE, one with $d_{xz}$ character and another one with $d_{yz}$ character.  We follow the LDOS at their energies as the bands evolve with increasing $\Delta_{\mathrm{OS}}$.  The LDOS patterns are outlined in red (green) at the energy of the BE of the $d_{xz}$ ($d_{yz}$) band.  The patterns outlined at $\omega=25,33$~meV ($\omega=63,85$~meV) for the $d_{xz}$ ($d_{yz}$) band are very similar to each other and distinct from the other patterns; When the energy of the band at the zone center shifts to a new value, there is a corresponding LDOS pattern that also shifts to this new value.  In contrast, patterns at other energies are not as sensitive to small OS changes, as seen by comparing the two patterns not outlined at the energies $\omega=25,40,$ or $85$~meV. We show this splitting effect at OS magnitudes of 30 and 60 meV, and it is robust down to the resolution limit of around 10 meV.  This leads to one clear signature of OS: As one increases OS, there are two LDOS patterns that increasingly shift away from each other, leaving the other patterns unchanged.  Measuring the energy separation between these two patterns yields the magnitude of OS. 

These patterns display a rotation relation that is unique to OS:  Applying a $90 \degree$ rotation about the $z$ axis to the LDOS pattern corresponding to the $d_{xz}$ BE yields the $d_{yz}$ pattern, because the orbitals themselves interchange under this rotation.  The similarity is however not perfect because of contributions from other orbitals and different bands at these energies, as explained below.  These two patterns superimpose at the same energy for $\Delta_{\mathrm{OS}}=0$, where the two BE are degenerate.  OS is therefore identified by two patterns at nearby energies that mostly map onto each other under 90$\degree$ rotation; these patterns superimpose at high temperatures where a putative OS is small.

Although QPI is the Fourier transform of the LDOS, it only exhibits part of the OS signatures and seems less suitable for tracking OS.  Fig. \ref{fig:OSmig_qpi}(a) shows the QPI patterns at the same energies and $\Delta_{\mathrm{OS}}$ values as in the outlined LDOS patterns in Fig. \ref{fig:OSmig_LDOS}.  The patterns at $\omega=63,85$~meV match, tracking the $d_{yz}$ BE as OS increases.  However, one does not obtain a $d_{xz}$ pattern shifting at lower energies.  There is thus no signature of two patterns that map onto each other under $90\degree$ rotation.  The reason is that the $d_{xz}$ pattern is obscured by contributions from other orbitals.  This is seen by projecting out these contributions in Fig. \ref{fig:OSmig_qpi}(b), plotting only the $d_{xz}$ and $d_{yz}$ contributions.  This recovers the full OS signature: Two QPI patterns shift away from each other with increasing OS and these two patterns map onto each other under $90 \degree$ rotation (apart from minor obscuring at $\omega=33$~meV due to other $d_{xz}$ contributions).  However, this projection is not currently feasible in experiments.  Conversely, in the total LDOS, contributions from other orbitals are concentrated within two sites of the impurity, but they are less significant at further distances where OS patterns appear.  Therefore, the LDOS seems to be a better probe than QPI for detecting and measuring OS.
% Why are contributions for LDOS far away?  As far as I know, they just work out that way for this band structure.

\begin{figure}  
\begin{center}
\includegraphics[width=\columnwidth, viewport=0 1.25in 8in 8in]{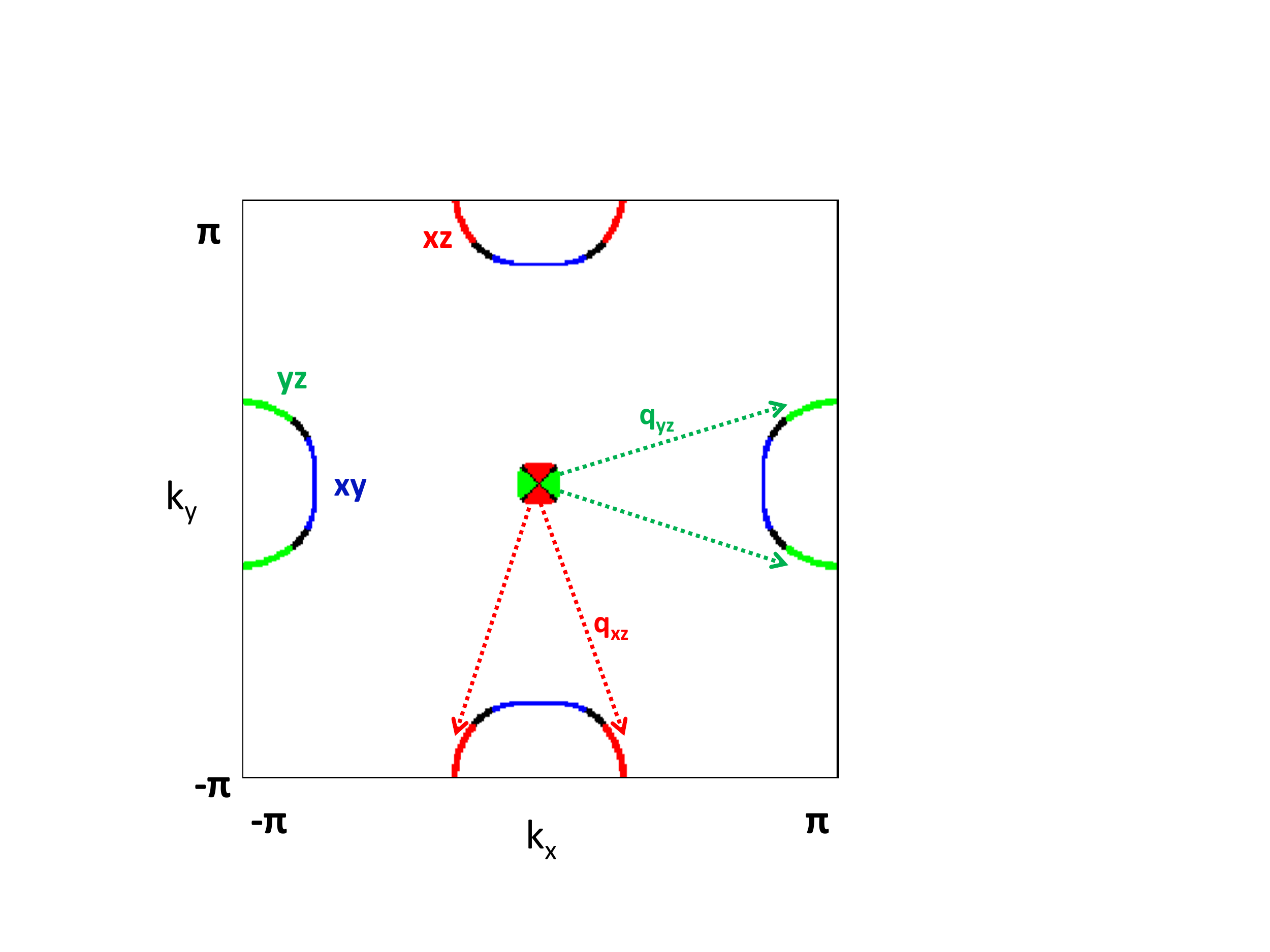}
\end{center}
\caption{(Color online) The CEC at the energy of the two BE in the $\Delta_{\mathrm{OS}}=0$ state.  The arrows are examples of $\mathbf q$ vectors for intra-orbital impurity scattering from the two high spectral weight BE.}
\label{fig:CECOS}
\end{figure}

The reason that LDOS tracks BE is understood by considering the LDOS's relation to the band structure.  A BE can cause a distinct STS pattern due to its large spectral weight in a small part of $\mathbf{k}$ space.  The resulting QPI pattern resembles the constant energy contour (CEC) of the bands at the energy of the BE, similar to the observed spectra in the superconducting state. \cite{Hanke2012}  The corresponding LDOS pattern follows the BE energy as it shifts.  This also explains the $90 \degree$ rotation relation: due to intra-orbital scattering, the BE only scatters to portions of the CEC of similar orbital content, as shown in Fig. \ref{fig:CECOS}.  Since the $d_{xz}$ and $d_{yz}$ portions follow the rotation relation, the same relation appears in the LDOS patterns.  In summary, OS can be identified and tracked by the temperature (or $\Delta_{\mathrm{OS}}$) dependence of two distinctive LDOS patterns, similar under $90\degree$ rotation, that split in energies corresponding to the BE.

The measurement of the OS magnitude and the comparison to the SDW gap may clarify the dependence of OS on magnetism.  We treated the case of large OS at the zone center, which is only predicted by theories such as ferroorbital ordering. \cite{Daghofer2012a,Daghofer2012}  In scenarios like OS driven by spin fluctuations, \cite{Fernandes2012b} where OS at the zone center is much smaller, STS may still measure it.  Also, OS is larger at other momenta, and BE at the corresponding energies would produce $90\degree$-rotated patterns.   Another possibility is that SDW causes OS as bands of similar orbitals repel each other.  This is noticeable in Fig. \ref{fig:bands}(b), where the $d_{yz}$ gapping shifts the upper band above the $d_{xz}$ band.  However, this repulsion can shift two bands apart by at most the SDW gap magnitude.  Moreover, we find the SDW-induced OS to be much smaller than the average SDW gap, by more than an order of magnitude at the zone center, for example (Fig. \ref{fig:bands}(b)).  This is also suggested in Fig. \ref{fig:mij}(b) by the small difference in orbital occupation, reaching a maximum of 0.02.  Therefore, observing OS of the same order as an SDW gap would imply that they are mostly independent of the SDW, and an even larger OS would definitively rule out SDW gaps as the sole cause.  Conversely, a small OS allows for a dependence on the SDW, and momentum dependence may further support the connection to spin fluctuations.
%theoretical limit of OS due to SDW is 2 \Delta, but this is complicated to rigorously define

\subsection {Signatures of a Collinear Spin-Density Wave}
\label{sec:SDWmigration}

In principle, any BE that shifts in energy leaves a signature in the LDOS.  Here, we present signatures of BE splitting due to an SDW gap using $H = H_{0} + H_{\mathrm{OS}} + H_{\mathrm{SDW}}$.  The inclusion of OS does not change the qualitative conclusions.  For the SDW itself these signatures may not provide new information, as the spatially uniform DOS is often sufficient to measure gaps.  However, to identify OS as in the previous section, one needs to distinguish it from other forms of BE splitting.  We show several differences for the SDW and highlight that pattern similarity under 90$\degree$ rotation is unique to OS.

\begin{figure}  
\begin{center}
\includegraphics[scale=0.6, viewport=0in 0.25in 7in 8in]{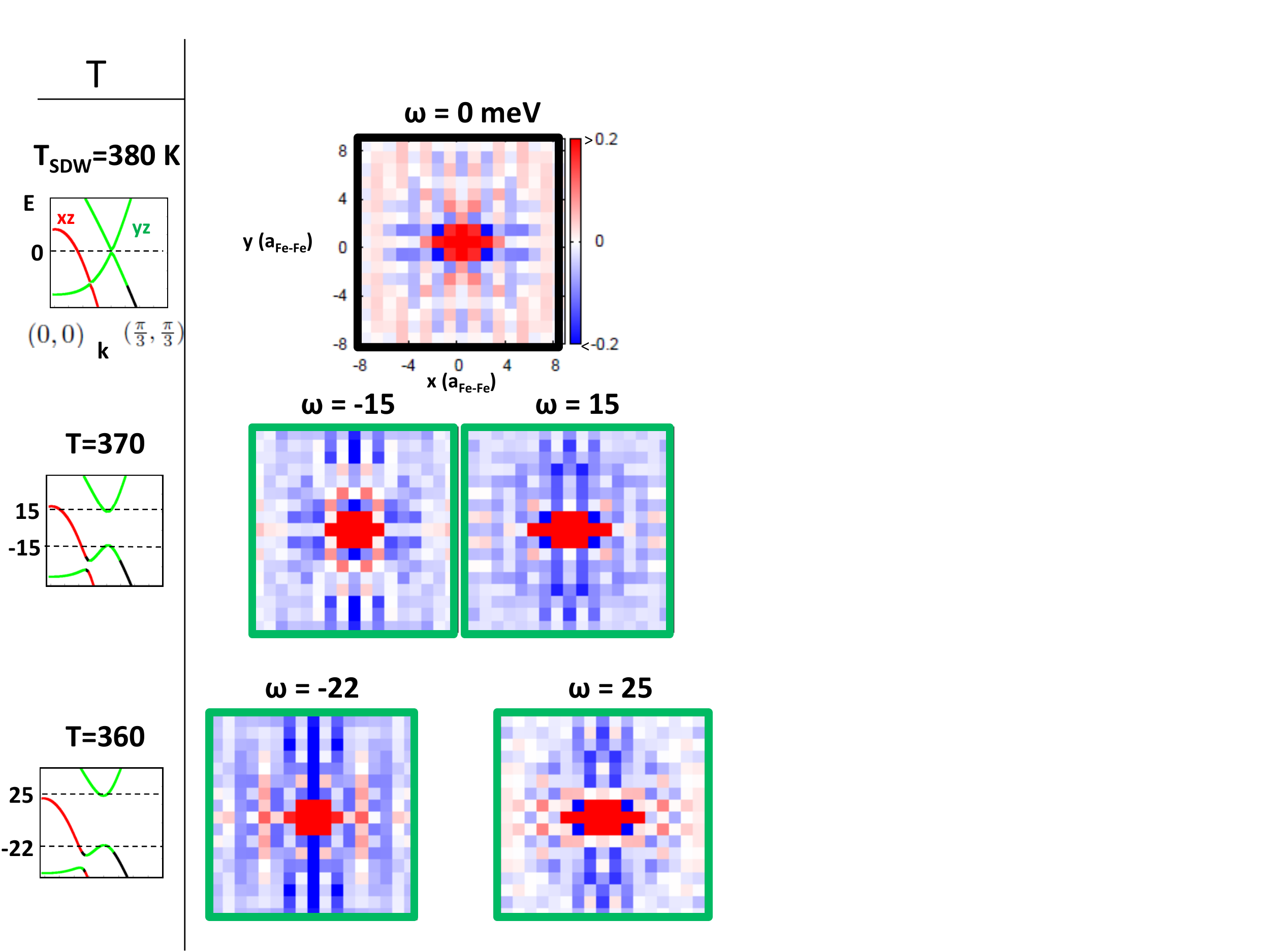}
\end{center}
\caption{(Color online) Signatures of the SDW in the impurity-induced LDOS, following similar conventions as in Fig. \ref{fig:OSmig_LDOS}. As the left column shows, each row displays results for decreasing temperature below $T_{\mathrm{SDW}}$ and hence increasing SDW band gap sizes.  The bands are in the MBZ and colored by their dominant orbital content. (For the black part of the bands, no orbital contributes more than $50\%$.)  Each dashed line tracks the energy of a gap edge as temperature varies.  At $T_{\mathrm{SDW}}$, the LDOS pattern shown is at the energy of the band crossing.  For lower temperatures the crossing is gapped, and patterns are at energies of the gap edges.  They are outlined in green as signatures of SDW gap formation.}
\label{fig:SDWmig_LDOS}
\end{figure}

\begin{figure*} 
\begin{center}
\includegraphics[scale=0.65, viewport=0in 0.25in 10in 8in]{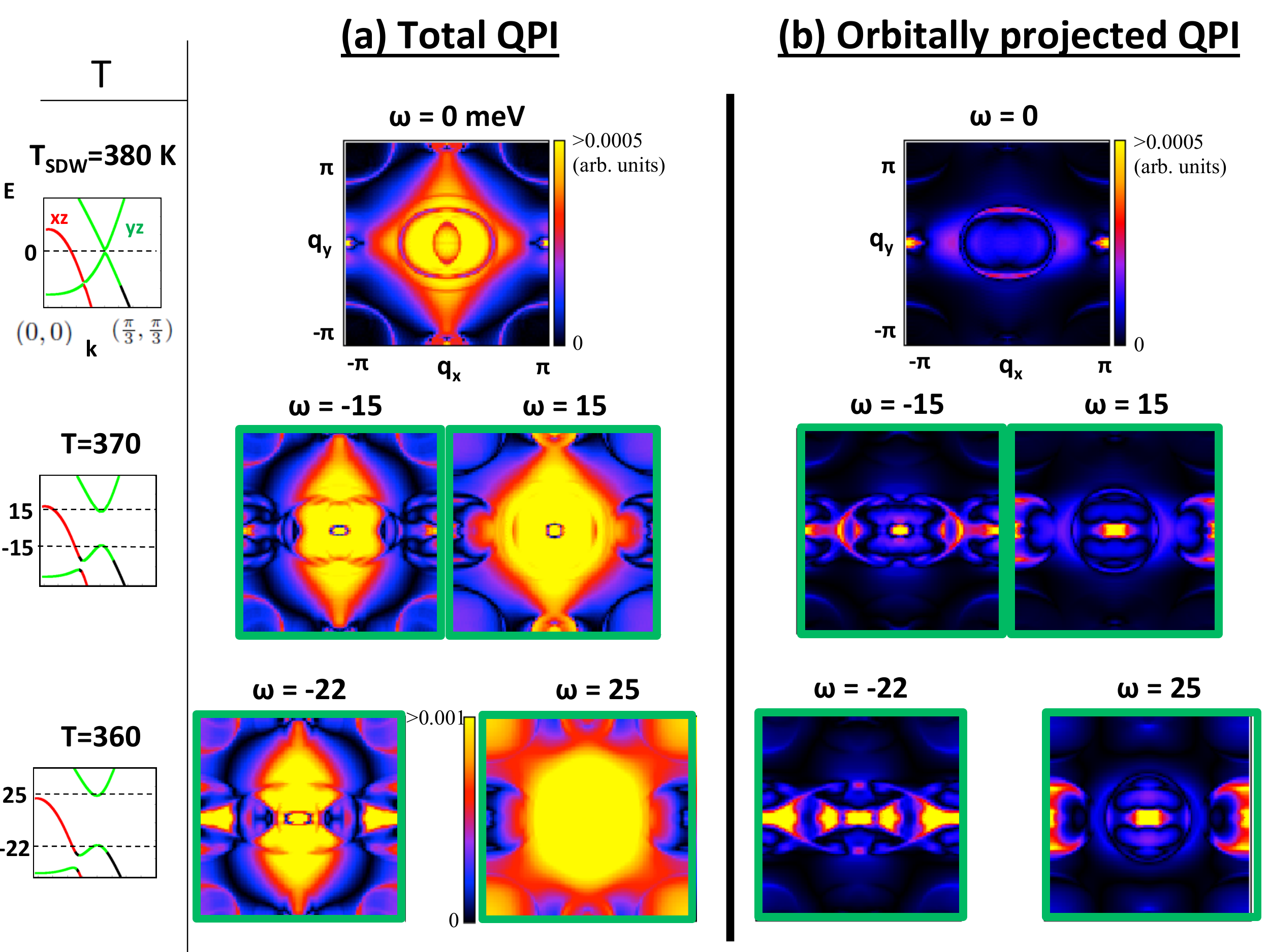}
\end{center}
\caption{(Color online) (a) Total QPI patterns (absolute value) for the same energies and OS as the LDOS in Fig. \ref{fig:SDWmig_LDOS}.  Here again, the SDW signatures are obscured.  All QPI patterns have the same intensity scale except at $\omega=25$~meV.  (b)  Orbitally projected QPI, showing only the $d_{yz}$ contribution, which displays SDW signatures.  All patterns have the same intensity scale.}
\label{fig:SDWmig_qpi}
\end{figure*}

The LDOS may track an SDW gap magnitude by BE splitting in a manner different from OS, as shown in Fig. \ref{fig:SDWmig_LDOS}.  As temperature decreases, an SDW band gap bordered by two BE grows.  At each temperature, we plot the LDOS at the energies of the BE.  As for OS alone, we find two patterns at $\omega=15,25$~meV ($\omega=-15,-22$~meV) that are similar.  For the SDW, these patterns track the gap magnitude.  However, several features are attributed specifically to the SDW rather than OS: The patterns of the two BE are similar without any rotation; a strong positive $\delta n$ is visible close to the impurity while a negative $\delta n$ extends away from it in the y direction.  This similarity is caused by BE on either gap edge having similar orbital content, since they originate from the same magnetic BZ (MBZ) crossing point.  Additionally, unlike OS for $\Delta_{\mathrm{OS}}=0$, there is no BE pattern at $T_{\mathrm{SDW}}$, since the gap closes and its BE disappear.  Furthermore, these signatures are not as clear as those of OS because the SDW reconstructs the bands, which alters the BE pattern as temperature decreases.  Also, there are more contributions from other bands in the energy range of the SDW gaps than the BE that orbitally split.  Note that these splitting effects are shown for SDW gaps of 30 and 60~meV and are robust down to the resolution limit of around 10~ meV.   Thus, although the LDOS patterns split for both SDW and OS, OS is distinguishable in several ways.

In QPI, these SDW signatures are again obscured by other orbitals, supporting the idea that BE signatures are easier to see in the LDOS rather than QPI.  The LDOS patterns are not obscured by other orbital contributions that are concentrated near the impurity.  Conversely, in Fig. \ref{fig:SDWmig_qpi}(a), the full QPI at the same energies and temperatures as the LDOS's in Fig. \ref{fig:SDWmig_LDOS} do not show SDW signatures.  Only when projecting out orbitals other than $d_{yz}$ (Fig. \ref{fig:SDWmig_qpi}(b)) are the two patterns at $\omega=15,25$~meV ($\omega=-15,-22$~meV) similar, with features encircling $(0,0)$ and $(\pi,0)$.  Unlike OS, all four patterns bear some similarity without rotation.  Again, this is understood by considering the CEC: Since both BE are of $d_{yz}$ content, they scatter to the same portions of the CEC, resulting in patterns similar without rotation.

% I don't want to say there is any evidence of intra-orbital gapping throughout the system, just that this particular gap we are looking at is clearly an intra-orbital gap.

\subsection{OS-SDW cooperation}
\label{sec:orbspin}

%Note that the degree to which symmetry is broken in QPI and LDOS patterns does not necessarily increase monotonically with increasing OS, since they follow the symmetry breaking in the CEC's.

Detecting and measuring OS magnitudes in the LDOS may be used to understand whether OS is driven by SDW order.  OS and the SDW appear to cooperate, since they both decrease in magnitude with doping.  Furthermore, theoretical studies have shown that OS enhances the magnetic moment. \cite{Lv2011} Our results also indicate that OS and SDW enhance each other, and there are LDOS features of BE splitting that can be attributed to this interdependence.

The emergence of SDW order in an orbitally split state increases the OS magnitude,\cite{Yi2011,Yi2012}  which is observable in STS.  We model again OS+SDW by $H = H_{0} + H_{\mathrm{OS}} + H_{\mathrm{SDW}}$ with $\Delta_{\mathrm{OS}}=60$~meV at $T=290$~K (vs. $T_{\mathrm{SDW}}=380$~K), like the OS determined experimentally at $T_{\mathrm{SDW}}$. \cite{Yi2011}  We find that the SDW increases the OS magnitude from 60~meV to 83~meV, observable in LDOS as an increase in the splitting between the BE patterns.  
%The LDOS in the case with SDW - not shown - does not show as clear OS signatures.

\begin{figure}  
\begin{center}
\includegraphics[scale=0.6, viewport=0in 1.5in 5.5in 7.5in]{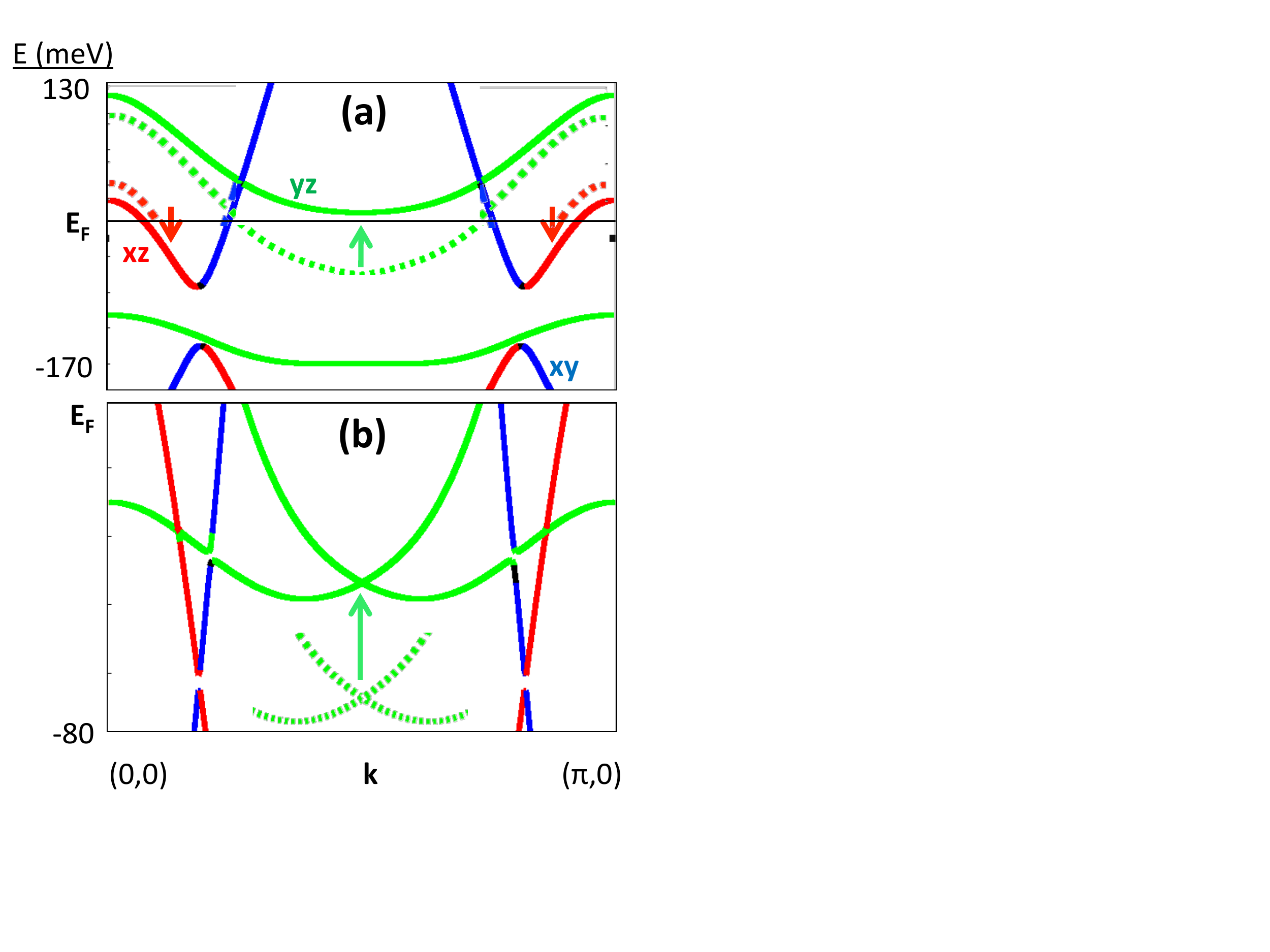}
\end{center}
\caption{(Color online) (a) Band structure for OS+SDW (solid lines) compared to some bands for OS only (dashed), folded into the MBZ.  The arrows highlight the effect on these bands' energies due to introducing the SDW. (b)  Bands for OS only in the MBZ (solid) compared to some bands without OS (dashed).  The arrow highlights the effects of introducing OS.}
\label{fig:EnhanceMP}
\end{figure}

This OS enhancement is understood by considering the bands; it is an example of a more general cooperation with an SDW.  Fig. \ref{fig:EnhanceMP}(a) compares the bands for OS only to those of OS+SDW.  It shows that the SDW moves the bands with $d_{yz}$ ($d_{xz}$) content near $\mathbf{k}=(0,0)$ to higher (lower) energies, leading to an increase of the OS magnitude observed in the STS patterns.  This occurs because the SDW involves bands of similar orbitals that repel each other (Sec. \ref{sec:theory}). This OS enhancement occurs throughout the BZ since the SDW-induced repulsion moves bands with $d_{yz}$ ($d_{xz}$) content across $E_F$ (arrows in Fig. \ref{fig:EnhanceMP}(a)), decreasing (increasing) its occupation.  This is verified by an increase in the orbital polarization per site, $n_{xz} - n_{yz}$, from 0.12 to 0.16.  

On the other hand, OS also enhances the SDW:  As highlighted by the arrow in Fig. \ref{fig:EnhanceMP}(b), adding OS to the state with no mean fields ($H=H_0$) shifts a $d_{yz}$ band crossing in the MBZ closer to $E_F$.   This improves the Fermi surface nesting and thereby SDW. \cite{Lv2011, Yi2012}  The zero-temperature magnetic moment increases from 0.43 for SDW only to 0.68 for OS+SDW with $\Delta_\mathrm{OS}=60$~meV.  In summary, there is two-way cooperation between OS and SDW throughout the BZ, as exemplified by the increased OS magnitude in the STS splitting.  This leaves the question whether OS not only cooperates but is even driven by the SDW, which LDOS signatures of OS may address.

%Proper explanation for Fig. \ref{fig:C2ness} ntbk2 P154f

\section{Summary and Experimental Realization}
\label{sec:conclusion}

In this work, we theoretically propose new experiments to identify features of each of OS and SDW and to clarify their interplay.   Our theory employs a realistic five-orbital model with a self-consistently determined mean-field SDW.  As temperature is decreased, two BE features in the LDOS are observed to shift away from each other, tracking OS or the SDW gap.  These features are easier to detect in the LDOS than in QPI, where they are obscured by contributions from other orbitals.  For OS, the features approximately map onto each other under $90\degree$ rotation, whereas SDW features are similar without rotation.  The energy difference between BE identified by STS patterns is the OS magnitude.  This addresses OS dependence on magnetism because a small OS at the zone center would support spin fluctuations or allow SDW gaps as the cause.  Conversely, an OS magnitude that exceeds gap magnitudes cannot be solely caused by SDW gaps.  Also, observing OS above $T_\mathrm{s}$ would address OS dependence on orthorhombicity.  Finally, considering SDW and OS together, we obtained their magneto-orbital cooperation as the SDW onset increases OS.

These results suggest further steps to discover OS signatures in experiments.  Unlike ARPES, STS can access energies above $E_F$, which may allow one to probe OS at the BZ center.  The model presented here employs a large OS for demonstration, but the existence and magnitude of OS in zone center bands above $E_F$ remains yet to be seen.  If that OS is much smaller than at the zone boundary, this may support nematic spin fluctuations as the mechanism, \cite{Daghofer2012a,Daghofer2012,Fernandes2012b} consistent with $s_\pm$- or $d$-wave superconductivity. \cite{Kuroki2008,Mazin2008b,Parish2008,Seo2008,Chen2009a,Wang2009}  Conversely, if the OS is not momentum dependent, that would support the ferroorbital ordering mechanism, consistent with $s_{++}$-wave superconductivity. \cite{Kontani2010, Yanagi2010a}

To experimentally search for OS in $dI/dV$ conductance measurements, the following steps seem appropriate.   A search within 100 mV above $E_F$ is promising because there is not much interference from other bands.  The signatures are easier to distinguish from background in a temperature sweep (Fig. \ref{fig:OSmig_LDOS}).  Since the OS is the main change in the bands for $T>T_{\mathrm{SDW}}$, this temperature range is most suitable to avoid interfering effects.  OS would appear as two 90$\degree$ rotated patterns at nearby voltages that split with decreasing temperature.  Further analysis may include an autocorrelation analysis and a subtraction between STS patterns at the same voltage for two nearby temperatures. Note that our model does not account for STS resolution issues and tunneling matrix elements.  However, since the proposed signatures are qualitative, they may be observable.

We have analyzed BE splitting in the iron pnictides, but the concept of LDOS and QPI tracking a BE energy can naturally be applied to other materials of interest.  Convenient materials would have fewer low energy bands to obscure the BE shifting, and BE at high symmetry points would present the clearest signatures.  The advantage is that many experimental techniques, including STS, still lack systematic means of identifying orbital degrees of freedom.  Therefore, our suggestion of observing OS in STS without the need for orbital resolution may be broadly applicable to transition metal oxide systems with partially filled orbitals.

\begin{acknowledgments}
We would like to acknowledge helpful discussions with C. Arguello, C.-C. Chen, M. Claassen, R. Fernandes, J. Kang, B. Moritz, E. Nowadnick, A. Pasupathy, E. Rosenthal, M. Sentef, R. Thomale, and M. Yi.  We acknowledge support from the U. S. Department of Energy, Office of Basic Energy Science, Division of Materials Science and Engineering, under Contract No. DE-AC02-76SF00515.  S. G. and A. P. K. acknowledge support by the DFG through TRR 80.
\end{acknowledgments}

\bibliographystyle{apsrev4-1}

\end{document}